\documentclass[letterpaper]{mc2025}

\usepackage{physics}
\usepackage{siunitx}
\usepackage{mhchem}
\usepackage{tikz}
\usepackage{textcomp}
\usepackage{subcaption}

\usetikzlibrary{shapes.geometric}
\usetikzlibrary{arrows.meta}
\usetikzlibrary{calc}
\usetikzlibrary{patterns}
\usetikzlibrary{fit}

\usepackage{nomencl} 
\makenomenclature

\DeclareSIUnit\pcm{pcm}
\DeclareSIUnit\gpm{gpm}

\title{Application of the Pathline Method to the Aircraft Reactor Experiment}

\addAuthor[mathis.caprais@cea.fr]{Mathis Caprais}{1}
\addAuthor{Nathan Greiner}{1}
\addAuthor{André Bergeron}{2}
\addAffiliation{1}{Université Paris-Saclay, CEA, Service d'\'{E}tudes des Réacteurs et de Mathématiques Appliquées, Gif-sur-Yvette, 91191, France}
\addAffiliation{2}{Université Paris-Saclay, CEA, Service de Thermo-hydraulique et de Mécanique des Fluides, Gif-sur-Yvette, 91191, France}
\Abstract{%
In this work, a new numerical method for the transport of Delayed Neutron Precursors (DNPs) is applied to the Aircraft Reactor Experiment (ARE). The pathline method is based on the Method of Characteristics (MOC) and leverages the pathlines of the liquid nuclear fuel to derive an integral form of the DNPs balance equation. The method has previously been tested on the CNRS benchmark and in a simplified 2D geometry where turbulent diffusivity was significant compared to advection. Here, the pathline method is applied to a real-world Molten Salt Reactor (MSR), the ARE. DNPs transport is implemented in the framework of the coupling between neutron transport solver APOLLO3\textregistered{} and computational fluid dynamics code TrioCFD, both developed at the French Atomic and Energy Commission (CEA). The DNPs concentration obtained with the pathline method were compared with those previously computed by TrioCFD, highlighting the importance of recirculation of fission products. The L-7 experiment was also replicated to demonstrate the method's capability.
}

\shortTitle{M&C 2025 Full Summary Template}

\authorHead{M. CAPRAIS}

\newcommand{\keff}{k_{\text{eff}}}

\DeclareSIUnit\pcm{pcm}

\begin{document}

\section{Introduction}

Molten Salt Reactors (MSRs) are liquid-fueled reactors that use molten salt as a fuel. Fissile materials are dissolved in the molten salt, which is often composed of light elements such as fluorine or chlorine. Historically, two MSRs have been built and operated, notably the Molten Salt Reactor Experiment (MSRE) at Oak Ridge National Laboratory (ORNL) in the 1960s \cite{macpherson1985molten}.

The earliest MSR project was the Aircraft Reactor Experiment (ARE), initiated by the US Air Force in the 1950s \cite{cottrell1955operation}. This reactor was designed as a compact, lightweight reactor for aircraft propulsion and used a mixture of highly-enriched \ce{LiF}-\ce{BeF2}-\ce{ZrF4}-\ce{UF4} as fuel. Operated at a power of \SI{2.5}{\mega\watt}, the reactor was moderated by beryllium oxide, and the fuel was circulated through the core by a pump. The reactor, which ran for 221 hours in 1954, was the first operational molten salt reactor and served as a precursor to the MSRE. Today, MSRs are being revisited for their potential in future nuclear energy production.

In MSRs, fission products are carried throughout the system by the circulating fuel salt. Among these, Delayed Neutron Precursors (DNPs) are particularly important because their neutrons can be emitted far from the original fission site, even out of the reactor core, therefore decreasing its reactivity. To account for the displacement of DNPs, existing neutronic codes must be upgraded or coupled with computational fluid dynamics (CFD) codes to model the transport of DNPs. This coupling can be computationally expensive, as the delayed neutron source might need to be recalculated at each power iteration.

In this work, the pathline method is applied in an existing coupling framework. This approach uses pathlines of the liquid fuel to derive an analytical form of the DNPs concentration. The development of the pathline method was motivated to improve the efficiency of the neutronics-thermal-hydraulics coupling between APOLLO3\textregistered{}/TrioCFD \cite{martin2024coupled, greiner2023new}. In the current APOLLO3\textregistered{} \cite{mosca2024apollo3}-TrioCFD \cite{angeli2015overview} coupling approach, the transport of DNPs is handled within TrioCFD by solving the time-dependent precursor transport equation until a steady state is reached, even when a steady-state solution is sought. This stationary-reaching temporal strategy is employed because TRUST/TrioCFD was developed for solving time-dependent problems, and does not offer specific methods for steady-state problems. By replacing this temporal scheme with a pathline-based stationary solution, the pathline method aims to provide a more efficient and optimized approach for DNPs transport. The ability of the pathline method to correctly calculate neutronics-thermal-hydraulic steady-states was demonstrated on the CNRS benchmark \cite{caprais2024newcalculationmethodusing} and extended to cases where turbulent diffusion is significant, opening up new possibilities for real reactor applications.

Here, the pathline method is applied to the ARE, with results compared to previous couplings in terms of accuracy and computational efficiency. A fuel flow experiment is also replicated to demonstrate the method's capabilities and the influence of precursors' boundary conditions.

This paper is organized as follows: the pathline method is reminded in Sec. \ref{sec:pathline} and the ARE is presented in Sec. \ref{sec:are} as well as the coupling strategy. Results are shown in Sec. \ref{sec:results}, followed by the conclusion in Sec. \ref{sec:conclusion}.
\section{Pathline Method}\label{sec:pathline}
In this section we recall the change of coordinates based on pathlines to obtain an analytical solution of the DNPs balance equation.
\subsection{DNPs Balance Equation}
The DNPs equation is a balance between production, decay, advection and diffusion. In its scaled form, it reads:
\begin{equation}
    \vb{u}^\prime\cdot\grad{C_j^\prime} - \frac{\mathcal{C}_j}{\mathcal{B}_j}\div{D^\prime\grad{C_j^\prime}} + \frac{1}{\mathcal{B}_j} C_j^\prime = \frac{S_j^\prime}{\mathcal{B}_j}, \quad +\,\mathrm{b.c}, \quad \mathcal{B}_j = \frac{u_0}{\lambda_j L}, \qq{and} \mathcal{C}_j = \frac{D_0}{\lambda_j L^2},
    \label{eq:dnps_eq}
\end{equation}
with \(C_j^\prime = C_j / C_0\) the scaled concentration of the \(j\)-th group of DNPs (\(C_0\), \SI{}{\per\cubic\meter}), \(\vb{u}^\prime\) the scaled velocity field of the fluid (\(u_0\), \SI{}{\meter\per\second}), \(D^\prime\) the scaled diffusivity coefficient of the DNPs (\(D_0\), \SI{}{\square\meter\per\second}) and \(S_j^\prime\) the scaled source term of the \(j\)-th group of DNPs (\(S_0 \propto \lambda_j C_0\), \SI{}{\per\cubic\meter\per\second}). The scaling factors \(\mathcal{B}_j\) and  \(\mathcal{C}_j\) represent the ratio of advection to decay and the ratio of diffusion over decay, respectively. From Eq. \eqref{eq:dnps_eq}, it is possible to see that long-lived DNPs are more affected by transport (advection and diffusion) as both the advection-reaction number \(\mathcal{B}_j\) and the diffusion-reaction number \(\mathcal{C}_j\) are proportional to their half-life.
\subsection{Analytical Solution along Pathlines}
The pathline method is hereafter briefly reminded, but as already been developed and tested \cite{caprais2024newcalculationmethodusing}.Assuming that the velocity field is known and divergence-free (\(\div{\vb{u}} = 0\)), it can always be written as the gradient of a potential vector \(\vb*{\Pi} = \grad{\varphi} + \chi \grad{\psi}\), so that the velocity field is \(\vb{u} = \grad{\chi}\cross \grad{\psi}\). The change of variables also requires the definition of pathlines, which are paths taken by tracers in the velocity vector field. These trajectories are defined by the kinematic equations of motions of the tracers, which are not coupled to the steady-state flow:
\begin{equation}
    \dv{\vb{r}}{\tau} = \vb{u} \qq{and} \tau = \int_{0}^{\ell} \frac{\dd{\ell^\prime}}{\abs{\vb{u}}},
    \label{eq:pathline_def}
\end{equation}
where \(\vb{r}\) is the position vector and \(\tau\), the time-of-flight (TOF) is the curvi-linear abscissa. The TOF is given by Eq. \eqref{eq:pathline_def} which is the time spent by a particle on a trajectory. Eq. \eqref{eq:pathline_def} implicitly assumes that the TOF at the beginning of the trajectory is equal to zero. Because each pathline is defined by unique values of \(\chi\) and \(\psi\) (\(\vb{u}\cdot\grad{\chi}=\vb{u}\cdot\grad{\psi}=0\)), these variables are dropped in the following. The triplet of variables \(\tau, \chi, \psi\) forms a new set of coordinates on which the DNPs equation can be rewritten. The diffusion term of Eq. \eqref{eq:dnps_eq} is moved to the right-hand side  to be treated as a source and the DNPs equation becomes:
\begin{equation}
    \dv{}{\tau}C_j^{(m)}\qty(\vb{r}\qty(\tau))  + \lambda_j C_j^{(m)}\qty(\vb{r}\qty(\tau))  = S_j^{(m-1)}\qty(\vb{r}\qty(\tau)), \qq{with} S_j^{(m-1)} \equiv S_j + S_{\mathrm{diff}}^{(m-1)} = S_j + \div{D\grad{C_j^{(m-1)}}},
    \label{eq:precursors_advection_pathlines_ode}
\end{equation}
where \(m\) is the iteration index. The source term \(S_j^{(m-1)}\) is updated at each iteration using the previous iteration's concentration field. The solution of Eq. \eqref{eq:precursors_advection_pathlines_ode} is given by:
\begin{equation}
    C_j^{(m)} \qty(\vb{r}\qty(\tau))
    = C_j^{(m)}\qty(\vb{r}\qty(\tau_{0}))e^{-\lambda_j\qty(\tau -\tau_{0})}
    + \int_{\tau_{0}}^{\tau}\dd{\tau^\prime} S_j^{(m-1)} \qty(\vb{r}\qty(\tau^\prime))e^{-\lambda_j\qty(\tau - \tau^\prime)} \qq{and} \tau_{0} < \tau.
    \label{eq:balance_precs_sol_incomp}
\end{equation}
Eq. \eqref{eq:balance_precs_sol_incomp} convoluates the modified DNPs source term of the previous iteration with an exponential decay along the pathline.
\subsubsection{Periodic boundary conditions}
Because the pathline method uses time-of-flight instead of the space variable, the implementation of periodic boundary conditions is straightforward if the out-of-core time is known. In the ARE, the fuel takes \SI{47}{\second} to traverse both the core and the heat exchanger before reentering the core \cite{bettis1957aircraft_ops}. Therefore, let \(\tau_c\) be the in-core time-of-flight and \(\tau_\ell\) be the out-of-core time-of-flight, the periodic boundary condition reads \(C_j^{(m)}\qty(\vb{r}\qty(\tau = 0)) = C_j^{(m)}\qty(\vb{r}\qty(\tau_c))\exp(-\lambda_j\tau_\ell)\), which in Eq. \eqref{eq:balance_precs_sol_incomp} leads to:
\begin{equation}
    C_j^{(m)} \qty(\vb{r}\qty(\tau)) = \frac{e^{-\lambda_j (\tau_l + \tau)}}{1 - e^{-\lambda_j (\tau_l + \tau_c)}}\int_{0}^{\tau_c}\dd{\tau^\prime} S_j^{(m-1)} \qty(\vb{r}\qty(\tau^\prime))e^{-\lambda_j\qty(\tau_c - \tau^\prime)} + \int_0^{\tau}\dd{\tau^\prime}S_j^{(m-1)} \qty(\vb{r}\qty(\tau^\prime))e^{-\lambda_j\qty(\tau - \tau^\prime)}.
    \label{eq:balance_precs_sol_periodic}
\end{equation}
Eq. \eqref{eq:balance_precs_sol_periodic} includes a finite integral of the source of the pathline, accounting for an infinite amount of recirculation as \( 1 / (1 - e^{-\lambda_j (\tau_l + \tau_c)})  = \sum_{k=0}^\infty e^{-\lambda_j (\tau_l + \tau_c)k}\).
\section{The Aircraft Reactor Experiment}\label{sec:are}
In this section, the ARE is described in terms of geometry, boundary conditions and material properties. Scaling numbers are also computed to assess the relative importance of precursors transport mechanisms.
\subsection{Description}
\begin{figure}[h]
    \centering
    \begin{subfigure}[b]{0.45\textwidth}
        \centering
        \includegraphics[width=1.1\textwidth, trim=18cm 0cm 18cm 0cm, clip]{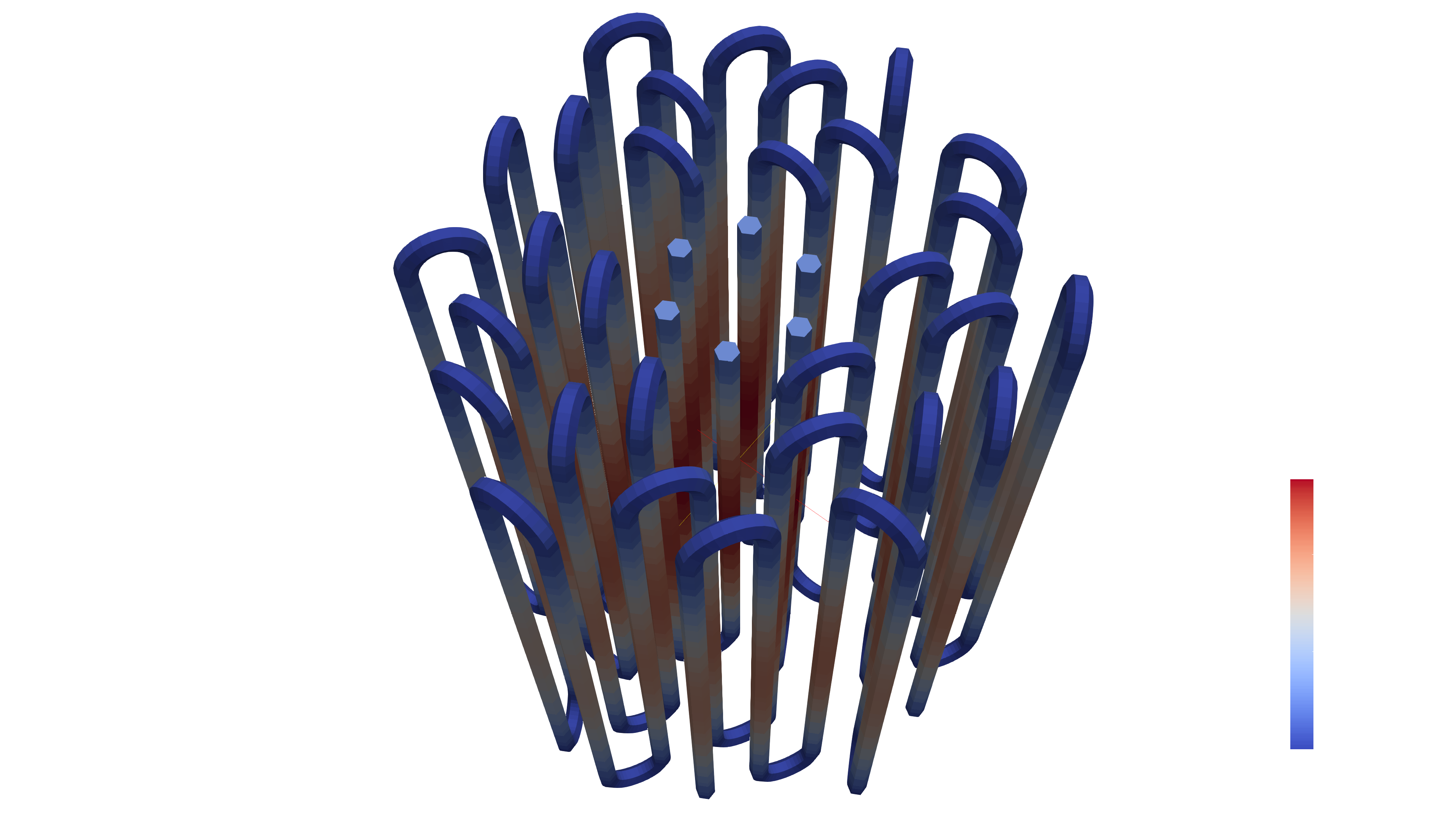}
        \caption{Fuel salt geometry in the ARE.}
        \label{fig:geometry}
    \end{subfigure}
    \hfill
    \begin{subfigure}[b]{0.45\textwidth}
        \centering
        \includegraphics[width=\textwidth]{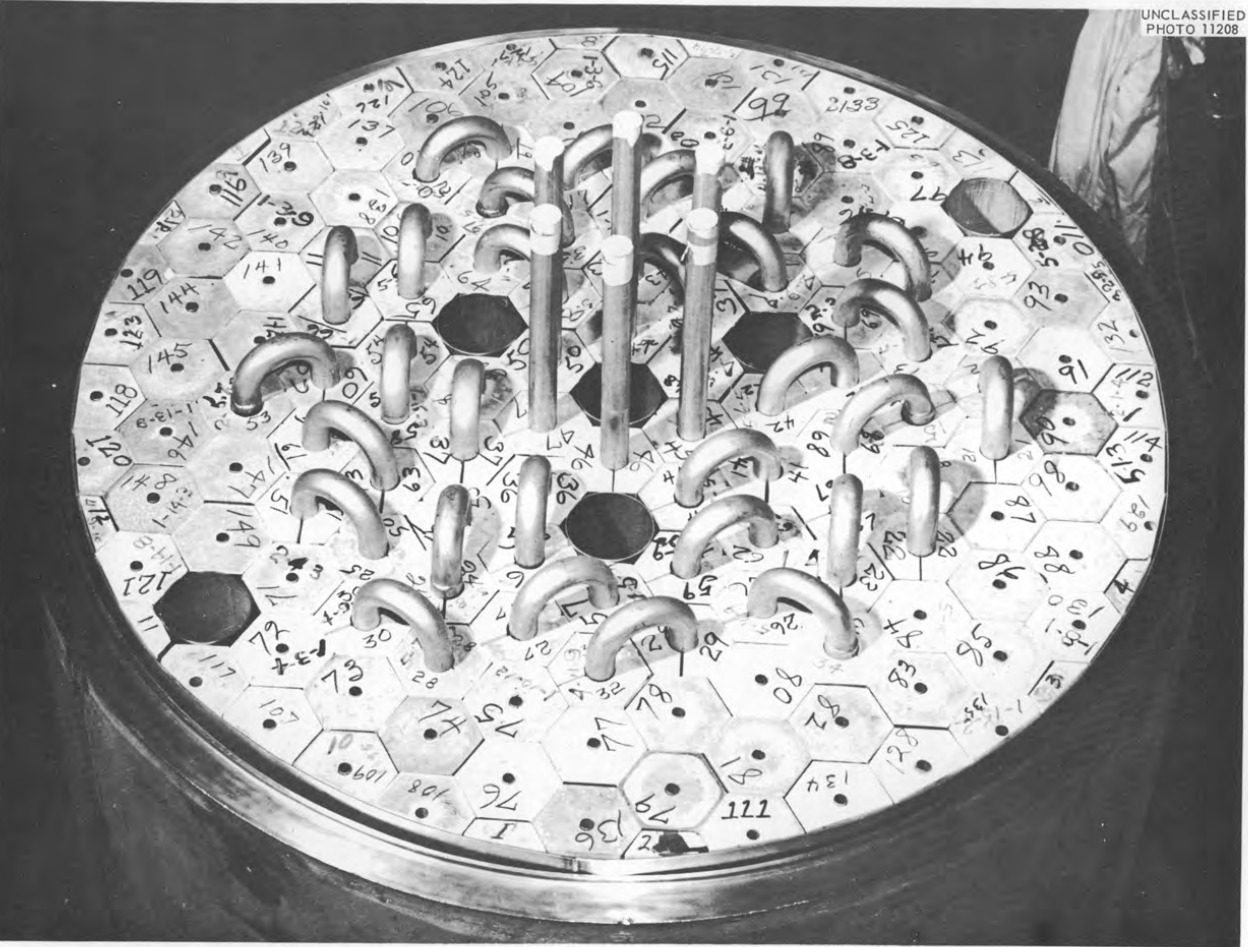}
        \caption{Original photograph of the ARE core and its six channels.}
        \label{fig:ARE}
    \end{subfigure}
    \caption{ARE fuel mesh and photograph of the core with the moderating beryllium oxide.}
    \label{fig:comparison}
\end{figure}

The ARE is a compact liquid fuel reactor. The nuclear fuel is a mixture of \ce{NaF}-\ce{ZrF4}-\ce{UF4} (\SI{52.8}{\percent}-\SI{40.73}{\percent}-\SI{6.18}{\percent}) highly enriched in \ce{^{235}U} (\SI{93.4}{\percent}). The fuel circuit is composed of six channels (Fig. \ref{fig:geometry}) passing through a beryllium oxide moderator cooled with liquid sodium, Fig. \ref{fig:ARE}. Each channel passes in the core eleven times before being cooled by a cross-current helium heat exchanger.
At a nominal volume flow rate of \SI{46}{\gpm}, the in-core time-of-flight computed with Eq. \eqref{eq:pathline_def} is \SI{15.36}{\second}, which is consistent with the ratio between the core volume and the flow rate of the fuel salt, \(V / q = \SI{17}{\second}\). As the total circulation time is \SI{47}{\second} \cite{bettis1957aircraft_ops}, the out of core time-of-flight \(\tau_\ell\) was adjusted accordingly.
\subsection{Scale Analysis}\label{sec:scale}
The scaling parameters for precursors physics (Eq. \eqref{eq:dnps_eq}) can be computed using values taken from different reports \cite{ergen1957aircraft_physics,bettis1957aircraft_ops,bettis1957aircraft_design,cohen1956physical}, and the standard eight-group delayed neutron precursors constants. As no value of mass diffusivity was available, a value was computed using the Einstein-Stokes diffusion coefficient \cite{di2022multiphysics}. The fission product radius was chosen to be \SI{1}{\pico\meter}, which is the lowest possible one and thus yielding the most penalizing (i.e. the largest possible) value of diffusivity. 
The scaling numbers are given in Table \ref{tab:scaling_numbers}.
\begin{table}[h]
    \centering
    \begin{tabular}{ccc}
    \hline
    Number & Symbol & Value \\
    \hline
    \hline
    Advection-reaction (avg.) & \(\mathcal{B}\) & \SI{8.2e-1}{} \\
    Advection-reaction (long-lived) & \(\mathcal{B}_1\) & \SI{5.05}{} \\
    Advection-reaction (short-lived) & \(\mathcal{B}_8\) & \SI{1.77e-2}{} \\
    Diffusion-reaction (avg.) & \(\mathcal{C}\) & \SI{2.7e-8}{} \\
    Diffusion-reaction (long-lived)  & \(\mathcal{C}_1\) & \SI{1.65e-7}{} \\
    Diffusion-reaction (short-lived) & \(\mathcal{C}_8\) & \SI{5.81e-10}{} \\
    \hline
    \end{tabular}
    \caption{Scaling numbers obtained at a temperature of \SI{1100}{\kelvin}.}
    \label{tab:scaling_numbers}
\end{table}
Long-lived precursors are much more affected by advection than short-lived precursors because they decay further away from their birth site due to a larger half-life. For all families of DNPs, the diffusion-reaction term is negligible compared to the advection-reaction term. Because the fuel flow is turbulent, turbulent diffusivity should also be considered. Away from the walls, \(\nu_t / \nu\) scales linearly with the Reynolds number \cite{pope2001turbulent}, which is of the order of \SI{1e4}{} in the ARE. Therefore, the values of the diffusion-reaction numbers are expected to be at least \SI{1e4}{} times larger, but still negligible compared to the advection-reaction numbers. Diffusivity will not be considered in the following.
\subsection{Coupled Neutronics and Thermal-Hydraulics}
\subsubsection{Meshes and Boundary Conditions}
The thermal-hydraulic problem is solved on three different domains: the fuel salt, sodium coolant, and the solid beryllium oxide (\ce{BeO}) moderator with the inconel pipes. Each of these domains is meshed separately. The neutron transport mesh differs from the thermal-hydraulic one, where the fuel salt bends are straightened for simplicity. The total fuel volume was conserved. Boundary conditions for the thermal-hydraulic simulation impose the fuel and sodium velocities and temperatures at the inlets, while adiabatic boundary conditions are applied at the core’s external boundaries. The projection of fields between the neutronic and thermal-hydraulic meshes is handled via the in-house CEA C3PO library, using the MEDCoupling (SALOME) library for P0-P0 field remapping. Each code begins by calculating cell-averaged values before projecting these values onto the other code's mesh.
\subsubsection{Coupling Strategy}
The neutronics and thermal-hydraulics codes are coupled using the fixed-point method. APOLLO3\textregistered{} calculates the neutron flux, power density field, and precursor source fields using an initial estimate of temperature. These fields are then passed to TrioCFD, which updates the velocity and temperature based on the thermal-hydraulic equations. TrioCFD sends the updated temperature fields back to APOLLO3\textregistered{}, cross-sections are updated and the steady-state problem is solved. The velocity is sent to a python module that applies the pathline method to obtain the DNPs concentration at each power iteration. The loop repeats until the exchanged fields converge to a specified tolerance. The coupling is managed through Python scripts that handle the remapping and synchronization of fields between the non-overlapping meshes of APOLLO3\textregistered{} and TrioCFD.
\subsubsection{Neutronic Calculations}\label{sec:neutronics}
To solve the neutron transport equation, APOLLO3\textregistered{} is employed with its \(S_N\) solver, MINARET. The homogenized cross-sections used in the core calculation are multi-parameterized. These cross-sections were computed during the lattice calculation on a 2D third of the core, with symmetry boundary conditions (the core is almost symmetric by a third). Macroscopic cross-sections were parameterized by the fuel temperature, the moderator temperature and the sodium temperature \cite{martin2024coupled}. One order of anisotropic scattering was considered and eight groups of delayed neutron precursors were used. Six energy groups are used, and vacuum boundary conditions are applied at the outer boundaries of the core.

In the original coupling, no recirculation of precursors was considered, meaning that no delayed neutron precursors re-enter the system once they leave. This is modeled by applying a zero-concentration boundary condition at the system inlet. APOLLO3\textregistered{} performs a steady-state neutronics calculation with initial temperatures. In the pathline method, cells crossed by pathlines are precomputed and times-of-flight within each cell are stored. The DNPs source is recalculated at each power iteration. The fuel mesh is an unstructured mesh composed of hexagonal prism elements, forming the six channels of the reactor with a total of \SI{4400}{} cells.
\subsubsection{Pathline Tracking}
The pathline method requires the pathlines of the liquid fuel to obtain the DNPs concentration within each mesh cell, Sec. \ref{sec:pathline}. 
Since the mesh of the fuel circuit is composed of consecutive hexagonal prisms and the velocity field coming from TrioCFD is cell-wise constant, the pathlines will consist in straight line segments, one in each prism, the exit point from one prism being the entry point in the next prism. The initial points are positioned at the centers of the inlet faces of the fuel circuit mesh, Fig \ref{fig:geometry}. In each hexagonal prism, the time-of-flight is then \(\Delta \ell / \norm{\vb{u}}\), with \(\Delta \ell\) the length of the pathline within the cell.
\subsubsection{Field Projection and Interpolation}
To handle the exchange of fields between the different meshes used by APOLLO3\textregistered{} and TrioCFD, the MEDCoupling library is used. Because the homogeneous cross-sections are multi-parameterized, Sec. \ref{sec:neutronics}, the temperature of the \ce{BeO} moderator around the fuel channels (Fig. \ref{fig:ARE}) is used to evaluate the cross-sections in the fuel salt using the temperatures of the surrounding regions.
\section{Results}\label{sec:results}
To verify our implementation, the concentrations obtained with the pathline method were first compared to those obtained with our previous implementation of the APOLLO3\textregistered{}/TrioCFD coupling \cite{martin2024coupled}, where TrioCFD solves the precursor transport with a stationary reaching temporal strategy. Then, the L-7 experiment of the ARE experimental program is replicated \cite{cottrell1955operation} with the pathline method. The performance of the pathline method is finally evaluated.
\subsection{Verification}\label{sec:verification}
\begin{figure}[h!]
    \centering
    \includegraphics[width=0.7\textwidth]{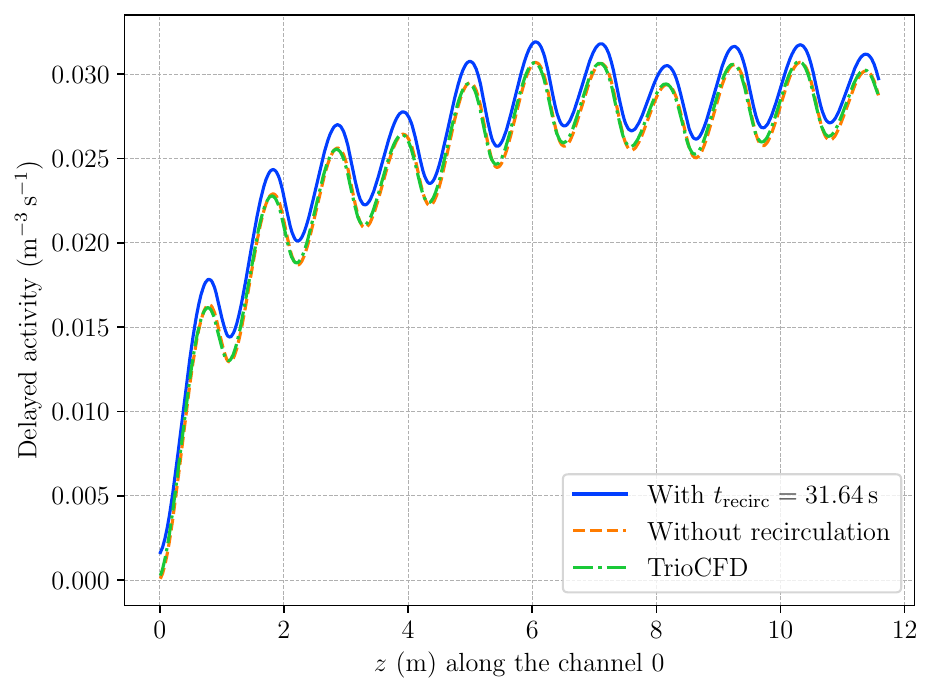}
    \caption{Total delayed activity in one of the fuel channels, computed with TrioCFD and with the pathline method (with and without recirculation).}
    \label{fig:verification}
\end{figure}

\begin{figure}[h!]
    \centering
    \includegraphics[width=0.7\textwidth]{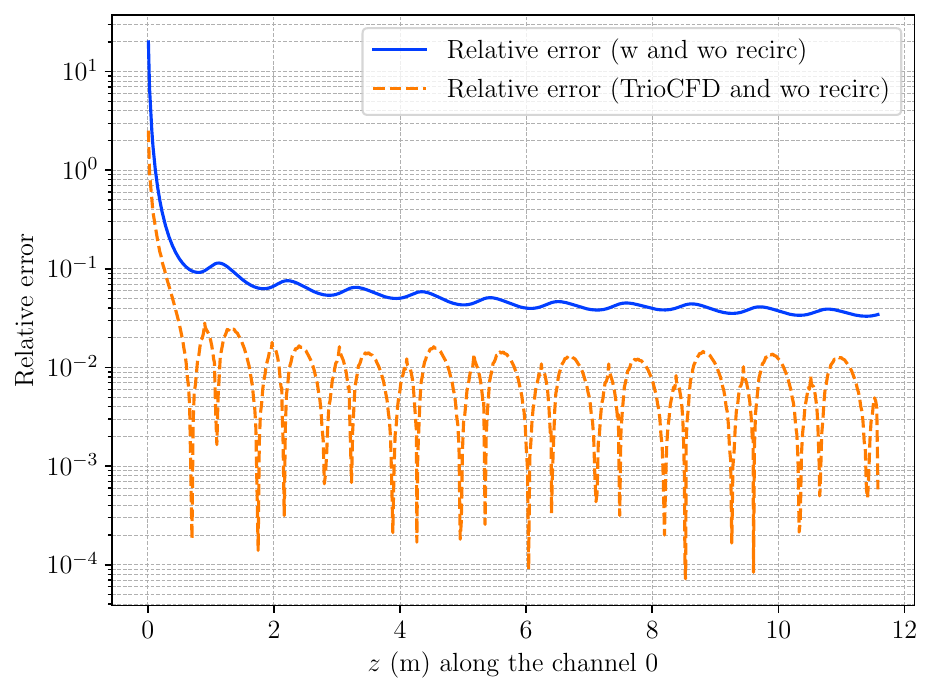}
    \caption{Relative error of the total delayed activity in one of the fuel channels, comparison between TrioCFD/pathline without recirculation and pathline with and without recirculation.}
    \label{fig:verification_error}
\end{figure}

To verify our implementation of the pathline method (without recirculation), the concentration of DNPs were first compared to those obtained with the original APOLLO3\textregistered{}/TrioCFD coupling. The coupling was performed with a thermal power of \SI{2.5}{\mega\watt} and a fuel flow rate of \SI{3e-3}{\cubic\meter\per\second} (nominal state) \cite{martin2024coupled}. The pathline method was called at the end of the coupled calculation and used the converged DNPs source calculated by APOLLO3\textregistered{}. Using the pathline method, the total delayed activity was computed and compared to that of the original coupling, Fig. \ref{fig:verification}.
The pathline method without recirculation exhibits a good agreement with the reference solution. The relative error in the channel is depicted in Fig. \ref{fig:verification_error}.
The relative error between the pathline method (without recirculation) and TrioCFD is high at the inlet of the fuel channel. This is possibly caused by how boundary conditions are handled in each code. Far from the inlet, the relative error is of the order of \SI{1}{\percent}. Finally, the total error on the delayed source was computed by integrating the DNPs delayed activity over the whole core and was \SI{2.61e-3}{} between TrioCFD and the pathline method without recirculation.
\subsection{Adding Recirculation}
Using the pathline method allowed an easy handling of the periodic boundary condition if the time spent of the core is known. Adding recirculation shifted the delayed activity by a constant, but the spatial profile remains unchanged, Fig. \ref{fig:verification}. This is due to long-lived precursors that have not completely decayed before re-entering the core. Accounting for DNPs recirculation increased the error. The relative error between the pathline method (with vs. without recirculation) is between \SI{100}{\percent} at the inlet and \SI{5}{\percent} within the channel. The relative error on the total delayed source between the pathline method (with recirculation) and TrioCFD was \SI{-4.59e-2}{}, which is coherent because the original implementation did not account for recirculation.
\subsection{L-7 experiment}\label{sec:l7}
\begin{figure}[h!]
    \centering
    \includegraphics[width=0.7\textwidth]{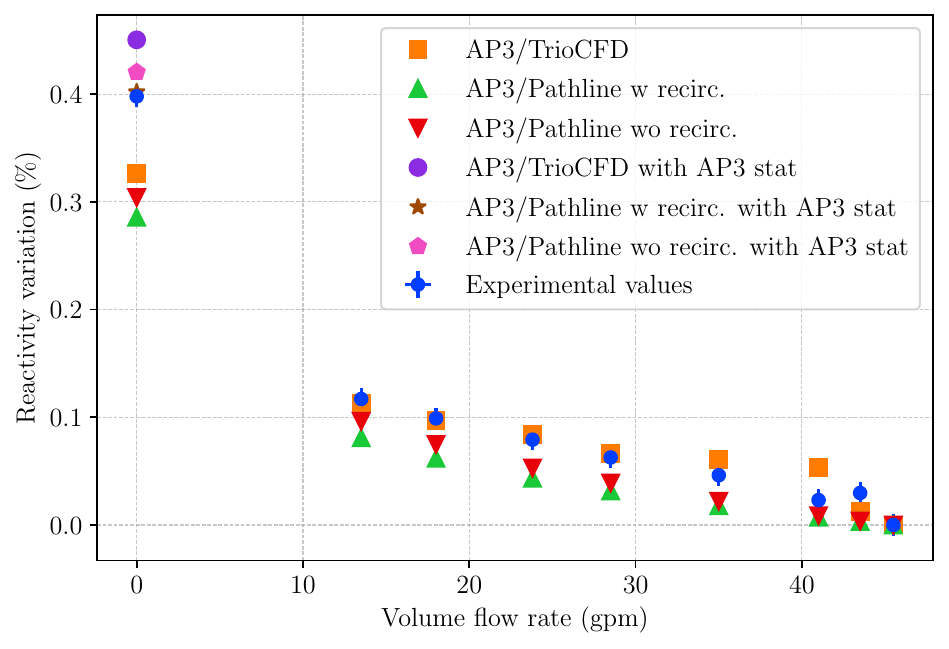}
    \caption{Reactivity variation in the L-7 experiment, computed with APOLLO3\textregistered{}, TrioCFD and the pathline method (with and without recirculation).}
    \label{fig:l7}
\end{figure}

In the L-7 experiment of the ARE experimental program, the reactor was brought to zero power after a fuel flow rate of \SI{46}{\gpm} was established. The fuel flow was then reduced down to \SI{0}{\gpm} and the position of the regulating rod was adjusted in order to compensate for the overreactivity introduced by more delayed neutrons being emitted in the core. The reactivity variation \(\Delta\keff / \keff\) was then estimated knowing the reactivity worth of the regulating rod. Numerically, the reactivity variation was estimated with the pathline method and APOLLO3\textregistered{} by changing the fuel flow rate and calculating \(\keff\). Because the recirculation times (time spent out of the core) \(\tau_\ell\) were not given at all flow rates in the report, they were inferred from the out-of-core time at the highest flow rate, \SI{46}{\gpm}. They were assumed to be inversely proportional to the flow rate, and therefore \(\tau_\ell \simeq \tau_\ell (46) Q_{46} / Q_{\text{gpm}}\) meaning that the out-of-core time increases as the flow rate decreases. Because temperatures of the reactor materials are not given in the report, they were assumed to be the ones of the L-8 experiment \SI{984}{\kelvin}, which took place minutes after the L-7 experiment. In Fig. \ref{fig:l7}, the reactivity variation at the highest volume flow rate (\SI{45.5}{\gpm}) was computed and compared to the experimental results.
The experimental results are given with their associated uncertainties. Because no uncertainty was available in the report, the uncertainty on the reactivity variation was inferred by the statistical dispersion of the reactivity variations measured at the same flow rate. The uncertainty on the flow rate was assumed to be \SI{0.5}{\gpm}, corresponding to the last digit of the flow rate measurements.

The values of \(\keff\) computed at zero volume flow rate with every method differed from the one computed by APOLLO3\textregistered{} only by at least \SI{100}{\pcm}. This is due to projection between meshes. As the bends of the fuel circuit are straightened in the neutronic mesh, the concentration of DNPs projected from the thermal hydraulic mesh (with bends) are lost.

While at intermediate flow rates a difference appears between the pathline method and the experimental results, the total reactivity variation of \SI{0.40\pm 0.01}{\percent} measured is remarkably consistent with the one obtained numerically with the value of \(\keff\) computed with APOLLO3\textregistered{} only and the pathline method with recirculation, \SI{0.40}{\percent} (uncertainties on the numerical reactivity variations are negligible compared to experimental values). This is to be compared with the one obtained with AP3TrioCFD, worth \SI{0.45}{\percent}. Without periodic boundary condition, the pathline method and TrioCFD overestimate the total reactivity variation as more DNPs are lost.
\subsection{Performance}
Both methods for DNPs calculation were run on one CPU. The python module is compiled with \texttt{numba} v.0.60. The time took by each method to solve the DNPs concentration was measured during a zero power calculation, Sec. \ref{sec:l7}. The initialization of TrioCFD took \SI{1.4}{\second}. For the pathline method, the initialization and tracking of pathlines within the geometry represent an overhead and accounts for most of the computational time, taking roughly \SI{2}{\second} when the velocity field is known.

Then, the time took by each method to compute the DNPs concentration with a source coming from the neutronic calculation was measured.  Averaging over the \SI{20}{} calls, TrioCFD takes \SI{1.36\pm0.01}{\second} (excluding the first call that takes \SI{21}{\second}) per call to calculate the DNPs concentration with the updated DNPs source. For the pathline method, excluding the first call, the average time spent calculating the DNPs concentrations was \SI{0.096\pm0.001}{\second}. The first call takes significantly longer, leading to a higher initial computation time due to the warm-up (just-in-time compilation) of the Python module that handles the pathline method, resulting in an initial time of \SI{0.7}{\second}. The number of power iterations is also different, while the AP3/TrioCFD coupling needs \SI{20}{} exchanges of field to converge, the AP3/Pathline method needs \SI{7}{}.

Therefore, the pathline method was one order of magnitude faster than solving the DNPs equations with TrioCFD once pathlines are traced. However, this should be balanced with the fact that the mesh of the ARE, while unstructured, is relatively simple.
\section{Conclusion}\label{sec:conclusion}
In this work, the pathline method has been applied to the ARE. The pathline method uses the pathlines of the liquid nuclear fuel to obtain an iterative solution of the DNPs equation, which is then used to estimate the DNPs concentration within each mesh cell. This method was already applied to academic problems, such as the CNRS benchmark or a simplified 2D reactor with turbulent diffusion of DNPs.

The method was first verified against a previous calculation of the DNPs concentration within fuel channels of the ARE. It was compared with a previous implementation of the coupling. After this verification, the implementation of periodic boundary conditions was tested and its influence of the total DNPs activity in the core was calculated. Then, the method was used to replicate the L-7 experiment of the ARE experimental program. The total reactivity variation between the highest flow rate and the lowest flow rate was computed and compared to the experimental results. While some discrepancies appeared because of the projection of fields between meshes, the results obtained with pathline method were consistent with the experimental results, with a reactivity variation of \SI{0.4}{\percent} to be compared to the experimental value of \SI{0.40\pm0.01}{\percent}. Excluding overheads, the pathline method was faster than the previous implementation of the coupling. Indeed, while TrioCFD solves an unsteady-state problem with time steps, the pathline method directly solves the steady-state problem.

The pathline method is therefore a promising tool for the calculation of DNPs concentration in liquid-fueled reactors, as it is both accurate and computationally efficient. However, even if the geometry considered in this work was more challenging than the previous academic problems (unstructured mesh), the fuel flow remained very simple by being represented on consecutive hexagonal prisms.

The next step will be to apply the method to more complex flow patterns and meshes, which can be found in pool-type reactors such as the MSFR. In this case, the tracking of pathlines will be more challenging, as the flow is not as regular as in the ARE.
\section*{Acknowledgments}
The authors are grateful to François Martin for his help with the APOLLO3\textregistered{}/TrioCFD coupling. APOLLO3\textregistered{} is a registered trademark of CEA. The authors gratefully acknowledge EDF and Framatome for their long-term partnership and their support.
\bibliographystyle{unsrt}
\bibliography{bib.bib}
\end{document}